\documentclass[twocolumn,amsmath,amssymb,floatfix,groupedaddress,prl]{revtex4}
\usepackage{graphicx}
\usepackage{color}
\usepackage{ulem}

\begin{document} 

\title{Electrostatically tunable optomechanical ``zipper'' cavity laser}

\author{R. Perahia}
\email{rperahia@caltech.edu}
\author{J. Cohen}
\author{S. Meenehan}
\author{T. P. Mayer Alegre}
\author{O. Painter}
\affiliation{Thomas J. Watson, Sr., Laboratory of Applied Physics, California Institute of Technology, Pasadena, CA 91125}
\date{\today}

\begin{abstract} 
A tunable nanoscale ``zipper'' laser cavity, formed from two doubly clamped photonic crystal nanobeams, is demonstrated.  Pulsed, room temperature, optically pumped lasing action at $\lambda = 1.3$~$\mu$m is observed for cavities formed in a thin membrane containing InAsP/GaInAsP quantum-wells.  Metal electrodes are deposited on the ends of the nanobeams to allow for micro-electro-mechanical actuation.  Electrostatic tuning and modulation of the laser wavelength is demonstrated at a rate of $0.25$~nm/V$^2$ and a frequency as high as $\nu_{m}=6.7$~MHz, respectively.
\end{abstract}

\maketitle

There has been growing interest in combining micro-optical resonators with micro-electro-mechanical systems (MEMS) to create tunable photonic elements such as lasers, couplers, and filters \cite{ref:Lee_MCM,ref:Huang_MCY,ref:loncar_dynamically_2009}. In particular, MEMS-tunable lasers are promising for applications such as on-chip spectroscopy \cite{ref:duarte_tunable_2008} and lightwave communication \cite{ref:Chang_C_J}, where large tunability and fast tuning speed are desirable.  Additionally, as lasers and other resonant elements shrink below the micron scale, fabrication imperfections at the nanometer level can lead to substantial variance in the operating wavelength.  The development of widely tunable optical cavities, at both the micro- and nano-scales, can benefit from recent advances in the field of cavity optomechanics in which radiation pressure forces are used to actuate and detect the motion of the cavity structure~\cite{ref:kippenberg_cavity_2008}. The strength of the radiation pressure force in such structures is quantified by an optomechanical coupling constant equal to the rate of change of the cavity resonance frequency $\omega$ with the amplitude of mechanical motion $x$, $g_{\text{OM}} = d \omega /dx$.  In the canonical example of a Fabry-Perot, the coupling constant is proportional to the inverse of the physical cavity length.  Recent theoretical and experimental work has shown that large gradient radiation pressure forces can be obtained in guided-wave nanostructures~\cite{ref:Povinelli_evanescent,ref:Tang_cavity,ref:eichenfield_picogram_2009}, with effective optomechanical cavity lengths less than or equal to the wavelength of light.

% Fig: Geometry 
\begin{figure}[ht!]
\begin{center}
\includegraphics[width=\columnwidth]{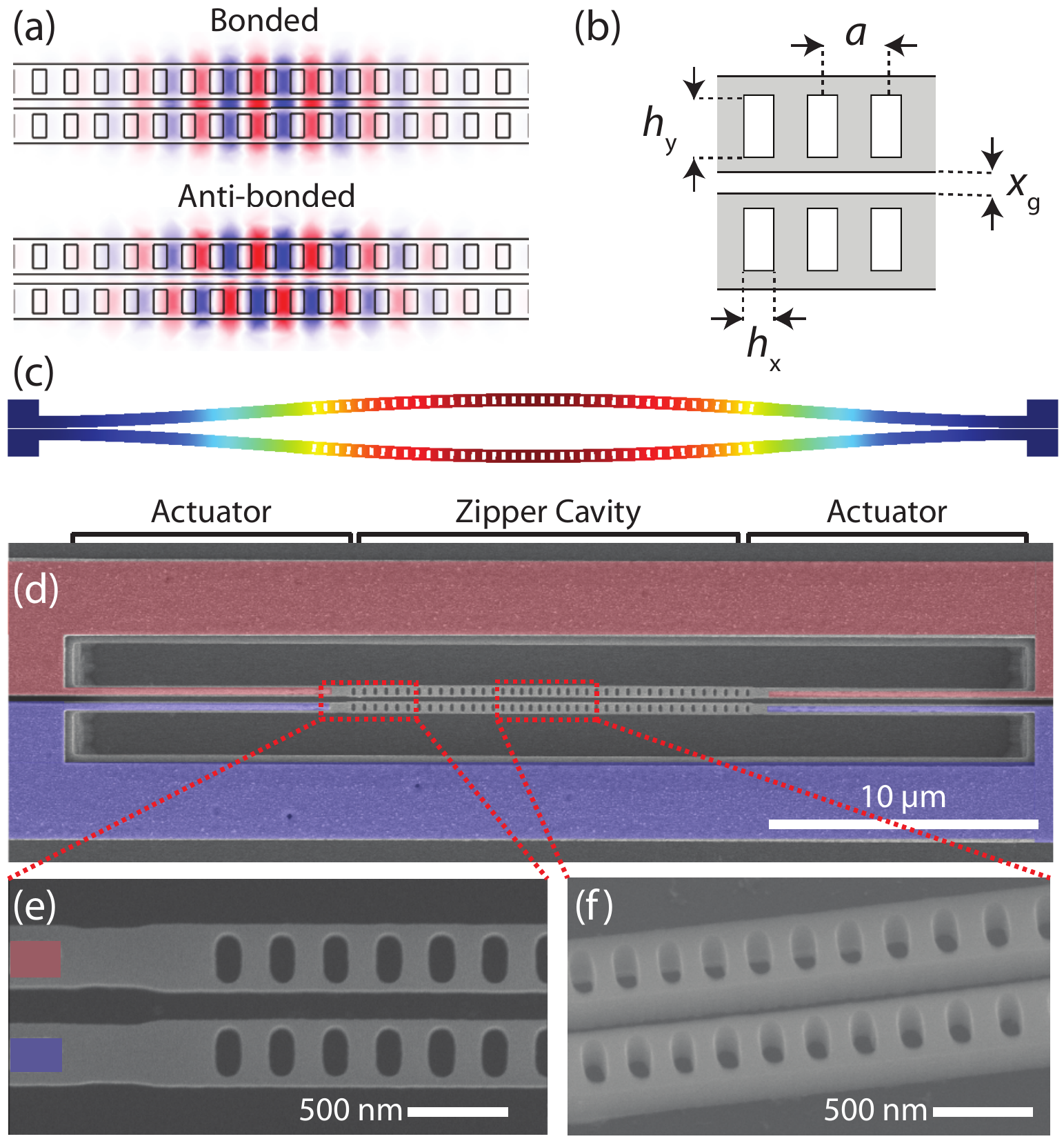}
\caption{(a) FEM simulated electric field of the fundamental symmetric and anti-symmetric zipper cavity modes. (b) Schematic of the zipper cavity.  The designed cavity structure has an inter-beam gap $x_g\approx 200$~nm, beam width $w=417$~nm, lattice constant $a=383$~nm, hole height $h_y=240$~nm and hole width $h_x=120$~nm. (c) FEM simulated fundamental in-plane mechanical mode ($\nu_{m,1}=1.66$~MHz, $m_{\text{eff}}=6$~pg, $k_{\text{eff}}=0.67$~N/m). (d) SEM micrograph of fabricated zipper cavity with MEMS actuators. Positive (red) and ground (blue) electrodes are highlighted. (e) Top view of the end-mirror section and (f) angled view of the central cavity region of the zipper cavity.}  
\label{fig:fig_1}
\end{center}
\end{figure}

Of particular interest in this work is the double nanobeam photonic crystal cavity, dubbed a zipper cavity~\cite{ref:eichenfield_picogram_2009,ref:Loncar_zipper}, in which giant radiation pressure effects have been measured.  In previous work~\cite{ref:Alegre_Perahia} we explored theoretically the optical, mechanical, and electrostatic properties of an integrated optomechanical and MEMS zipper cavity laser, comparing all-optical radiation pressure and electrostatic actuation for laser wavelength tuning and modulation.  Due to the small motional mass ($m_{\text{eff}}\sim$~pg) and large optomechanical coupling ($g_{\text{OM}}\sim100$~GHz/nm) of the zipper cavity, wavelength tuning rates approaching $\delta\lambda/\delta x\sim 1$~nm/nm at bandwidths well over a MHz were predicted.  Here we demonstrate such an integrated zipper cavity laser structure, formed in the InGaAsP semiconductor material system and operating in the $1300$~nm wavelength band.

As shown in Fig.~\ref{fig:fig_1}, the zipper cavity consists of a pair of nanobeams which are patterned with a linear array of holes and placed in the near-field of each other.  A small chirp in the period between holes near the center of the beams is used to form localized resonant modes~\cite{ref:chan_optical_2009}.  The predominantly transverse-electric (TE) optical modes of the two beams strongly couple and split into symmetric and anti-symmetric pairs (see Fig.~\ref{fig:fig_1}(a)), with the symmetric modes containing a large fraction of energy in the gap between the beams.  The localized optical modes are strongly coupled to the in-plane mechanical motion of the beams (Fig.~\ref{fig:fig_1}(c)), with symmetric modes tuning red and anti-symmetric modes tuning blue with a reduction in the inter-beam gap. 

  % The zipper cavity, shown schematically in comprised of two doubly clamped nanobeams patterned with a PC cavity and have a highly dispersive dependence on the gap between the nanobeams \cite{ref:eichenfield_picogram_2009}.  Optical defect modes are formed by a parabolic variation of the lattice constant at the center of the cavity. Using optical, mechanical, and electrostatic finite-element-method (FEM) simulations \cite{ref:chan_optical_2009,ref:Alegre_Perahia} we have designed zipper cavities appropriate for InAsP/GaInAsP strained quantum well laser material with photoluminescence centered at $\lambda = 1.35$~$\mu$m. As shown in Fig. \ref{fig:fig_1}(a) zipper cavities support optical modes that are classified by the symmetry of the electric field across the gap: symmetric (bonded) and anti-symmetric (anti-bonded).  With the reduction of nanobeam gap bonded modes (+) tune red and anti-bonded modes (-) tune blue.

% Fig: Experimental Setup
\begin{figure}[t!]
\begin{center}
\includegraphics[width=\columnwidth]{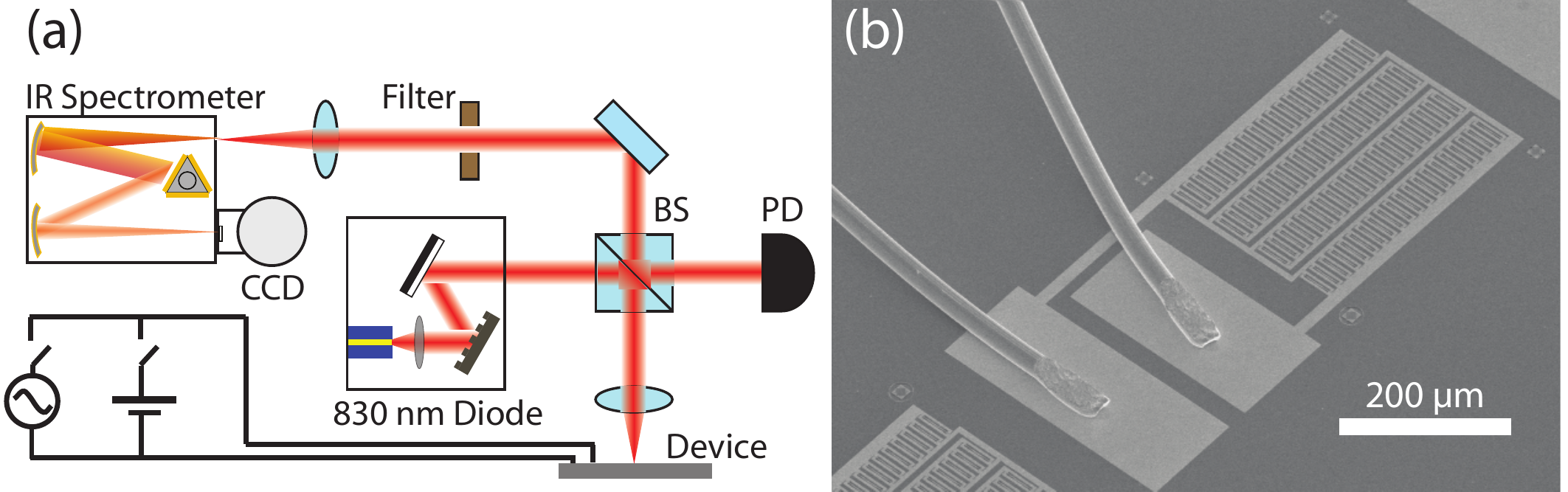}
\caption{(a) Micro-photoluminescence setup with static (DC) and modulation (AC) actuation circuits. (b) SEM micrograph of a wire-bonded device array.  In order to test a larger number of devices simultaneously, an array of $80$ zipper cavity lasers are connected in parallel to a common pair of contacts.} 
\label{fig:fig_4}
\end{center}
\end{figure}  

% The mechanical response of a zipper cavity is tailored by adding extensions to either side of the nanobeam structure \cite{ref:camacho_characterization_2009}. A $1$st order in-plane mechanical mode, which is most coupled to the $1$st order optical mode, is shown in Fig. \ref{fig:fig_1}(c). The nanobeam gap is capacitively actuated using insulated metal electrodes fabricated on top of the nanobeam extensions as shown in Fig. \ref{fig:fig_1}(e). 

As detailed in Ref.~\cite{ref:Alegre_Perahia}, the zipper cavity geometry naturally lends itself to integration with capacitive electro-mechanical actuators.  The zipper cavity structure studied here is fabricated from a $252$~nm thick InAsP/GaInAsP multi-quantum-well (MQW) layer grown on an InP substrate\cite{ref:Hwang2}.  Fabrication of the laser cavity begins with the deposition of an insulating thin $50$~nm layer of silicon nitride (SiN$_x$), followed by deposition and lift-off patterning of Ti/Au ($=10/190$~nm thick) metal electrodes.  A second SiN$_x$ masking layer is subsequently deposited, and an aligned electron beam lithography step is used to pattern the zipper cavity nanobeams over the metal electrodes.  An inductively-coupled plasma is used to etch the nanobeam pattern through the top InAsP/GaInAsP MQW layer.  Finally, the zipper cavity nanobeams are released from the substrate using a hydrochloric acid wet etch~\cite{ref:Perahia_R} that selectively removes the underlying InP substrate, followed by critical point drying to avoid collapse and stiction of the beams.  Scanning electron microscope (SEM) micrographs of a fabricated device are shown in Figs. \ref{fig:fig_1}(d-f).

The devices are tested using a micro-photoluminescence (micro-PL) setup shown schematically in Fig. \ref{fig:fig_4}(a).  The devices are optically pumped with an $830~$nm wavelength pulsed diode laser (pulse width $\delta T=20$~ns, pulse period $T=4~\mu$s).  A high numerical aperature ($0.4$), long-working distance ($32$~mm) objective lens is used to focus the pump beam on the sample surface with a spot size of approximately $2~\mu$m.  PL from the sample is collected through the same objective lens, separated from the pump using a beamsplitter/interference filter combination, and then dispersed and detected by an imaging spectrometer with an attached cooled InGaAs CCD array.  A known fraction of the pump power is picked off by the beamsplitter and sent to a calibrated detector to estimate the power incident on the device.  

% Fig: Laser data
\begin{figure}[b!]
\begin{center}
\includegraphics[width=\columnwidth]{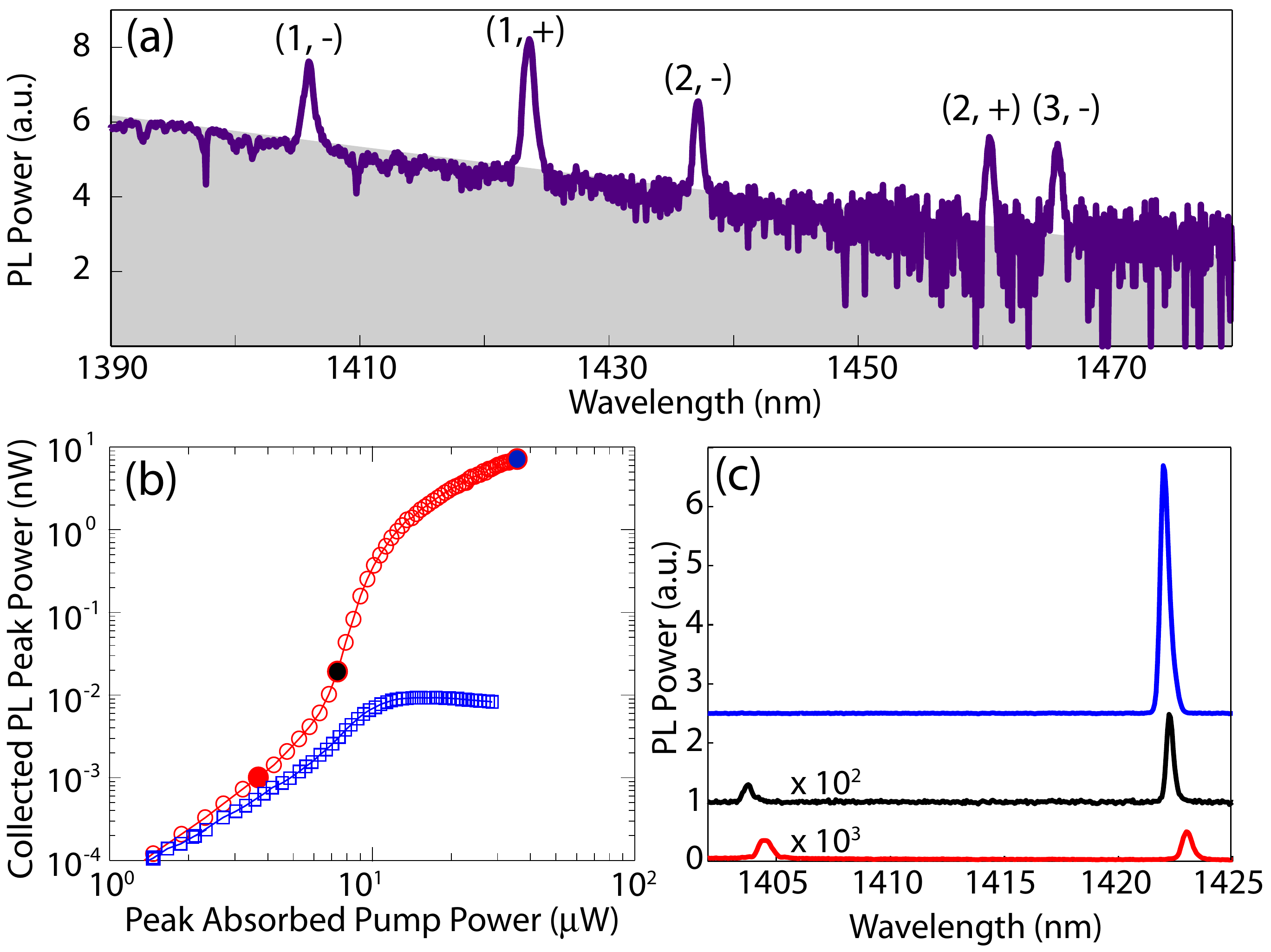}
\caption{(a) Sub-threshold spectrum of an optically pumped zipper cavity laser.  (b) Light-in versus light-out (LL) curve for the fundamental symmetric ($\circ$) and anti-symmetric ($\square$) zipper cavity modes. The peak absorbed pump power is estimated from the pump duty cycle ($1.1\%$), the fraction of the pump beam intercepted by the zipper cavity ($10\%$), and the material absorption ($10\%$).  (c) PL spectrum below (bottom curve), at (middle curve), and above (top curve) threshold corresponding to the filled circle data points in the LL curve of (b).} 
\label{fig:fig_2}
\end{center}
\end{figure}

A typical sub-threshold PL spectrum from a zipper cavity is shown in Fig.~\ref{fig:fig_2}(a).  Several longtitudinal orders of symmetric and anti-symmetric cavity mode pairs are visible in the spectrum, with the lower order modes occuring at shorter wavelengths and the higher order modes at longer wavelengths.  As indicated by the labelling, for each pair of modes the shorter wavelength mode is anti-symmetric and the longer wavelength mode is symmetric, as confirmed by their direction of tuning (see below).  A plot of the peak emitted power into the fundamental symmetric and anti-symmetric modes versus peak absorbed pump power is shown in Fig. \ref{fig:fig_2}(b).  A clear lasing threshold at an absorbed peak pump power of $10$~$\mu$W is evident for the symmetric $(1,+)$ mode, while the emission into the anti-symmetric $(1,-)$ mode saturates above threshold due to gain (carrier) clamping. Spectra of the laser emission below, at, and above threshold are shown in Fig. \ref{fig:fig_2}(c).  The laser linewidth narrows from $\delta\lambda\approx 0.7$~nm well below threshold to $\delta\lambda\approx 0.3$~nm at threshold, and then slowly rises to $\delta\lambda\approx 0.5$~nm well above threshold.  This anamolous linewidth behavior is not atypical of micro- and nano-scale semiconductor lasers~\cite{ref:Bjork4}; however, one possible explanation relates to the coupling of spontaneous emission fluctuations to the laser wavelength through radiation pressure.  Future experiments will seek to explore this possibility.

% The thermal Brownian motion contribution to the linewidth for the given $g_{\text{OM}}$ and spring constant $k_{\text{eff}}=0.7$ is estimated to be $\delta \lambda = 32$~pm at room temperature \cite{ref:rosenberg_static_2009}. The laser linewidth is therefore not limited by thermal mechanical fluctuations.

% Fig: MEMS 
\begin{figure}[t!]
\begin{center}
\includegraphics[width=\columnwidth]{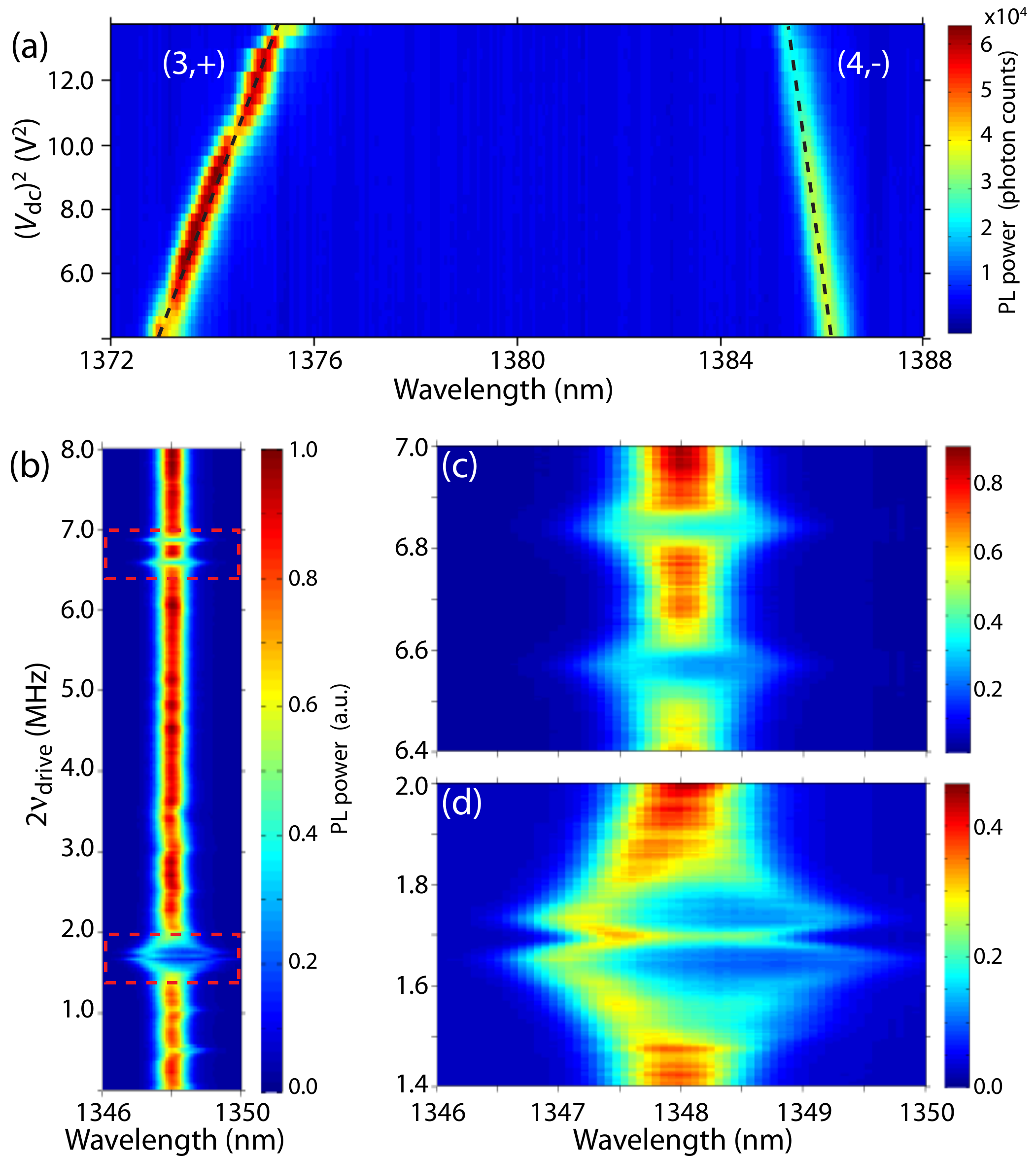}
\caption{(a) PL spectra of $(3,+)$ symmetric and $(4,-)$ anti-symmetric modes as a function of $(V_{dc})^2$.  Linear fits to the tuning curve are shown as a black dotted line. (b)  PL spectra of a zipper cavity laser versus MEMS drive frequency. Zoom in of the (c) third-order and (d) first-order in-plane mechanical resonances.} 
\label{fig:fig_3}
\end{center}
\end{figure}

The wavelength tuning properties of the zipper cavity laser is tested in two different ways.  First, measurements of the static tuning properties of the laser wavelength are measured by applying a variable DC voltage bias ($V_{dc}$) to the capacitive electrodes.  The spectra from a laser device with large static tuning, optically pumped sub-threshold to follow both symmetric and anti-symmetric modes, are shown in Fig.~\ref{fig:fig_3}(a).  Both modes are seen to tune linearly with the square of the applied voltage and in a direction corresponding to the reduction of inter-beam gap with increasing voltage, as expected for the capacitor geometry.  The $(3,+)$ symmetric mode tunes red at a rate of $0.25$~nm/V$^2$, for a total tuning of $\Delta \lambda \approx 3.5$~nm at an applied voltage of $V_{dc}=3.7$~V, whereas the $(4,-)$ anti-symmetric mode tunes in the opposite direction at a smaller rate of $-0.092$~nm/V$^2$.  The estimated inter-beam gap from the measured symmetric/anti-symmetric mode-splitting of this device is $x_{g}=240$~nm (consistent with SEM measurements), which from FEM simulation yields an optomechanical coupling constant of $g^{(3)}_{+}/{2\pi}=-27$~GHz/nm~($\delta\lambda/\delta x_{g}=0.16$~nm/nm) and $g^{(4)}_{-}/{2\pi}=11$~GHz/nm~($\delta\lambda/\delta x_{g}=-0.065$~nm/nm) for the $(3,+)$ and $(4,-)$ modes, respectively.  From these coupling constants a shift in gap size of $\delta x^*_{g}=21.7$~nm is inferred for an applied $V_{dc}=3.7$~V, which is in good agreement with the numerically modelled MEMS actuated gap shift of $\delta x^{**}_{g}=19.7$~nm.     

A second set of tuning measurements are performed by applying a small signal voltage modulation ($\delta V_{ac}$) to the capacitive electrodes over a drive frequency range from $\nu_{\text{drive}}=100$~kHz to $4$~MHz.  The optical response of the zipper cavity to the modulated voltage input is measured by recording the time-integrated PL spectrum.  The recorded spectra versus twice the drive frequency (the capacitive force scales as $(\delta V_{ac})^2$ resulting in mechanical actuation at $2\nu_{\text{drive}}$) of a laser device with a fundamental symmetric laser mode are shown in Fig.~\ref{fig:fig_3}(b).  Two resonances around $2\nu_{\text{drive}}=1.69$~MHz and $2\nu_{\text{drive}}=6.70$~MHz can be clearly seen in the modulation spectrum.  A zoom-in around each resonance is shown in Figs.~\ref{fig:fig_3}(c) and (d), indicating that each resonance is split into two resonance peaks.  FEM mechanical simulations of the fabricated zipper cavities (nanobeam length $L=31~\mu$m) yield mechanical resonance frequencies of $\nu_{m,1}=1.66$~MHz and $\nu_{m,3}=6.55$~MHz for the optomechanically coupled first and third order differential in-plane mechanical modes, respectively, in good correspondence with the measured resonances. Note that the second order in-plane mode is decoupled from the laser field due to its odd symmetry.  The presence of two resonance peaks around each mechanical mode frequency is indicative of slight asymmetries in the two nanobeams, resulting in independent beam motion. From the frequency and linewidth ($\delta\nu_{m,1} \approx 150$~kHz, due to squeeze-film damping~\cite{ref:rosenberg_static_2009}) of the first order mechanical resonance, the $3$-dB bandwidth and settling time for wavelength tuning of this device are estimated to be $3$~MHz and $1$~$\mu$s, respectively, comparable to recently demonstrated MEMS-tunable VCSELs \cite{ref:Huang_MCY}.        

The demonstrated static wavelength tuning ($0.25$~nm/V$^2$) and wavelength modulation rate ($6.7$~MHz) of the zipper cavity laser can be significantly improved with a few modifications to the cavity design.  A reduction in the nanobeam length of the zipper cavities to $L\approx5$~$\mu$m should be possible, enabling a fundamental in-plane mechanical resonance frequency approaching $50$~MHz, greatly enhancing the laser frequency modulation speed of the device.  Reduction in the inter-beam gap below $50$~nm, as has been demonstrated in passive zipper cavity geometries~\cite{ref:camacho_characterization_2009}, increases the optomechanical coupling and static wavelength tuning rate of the zipper cavity modes by more than an order of magnitude.  An increase in $g_{\text{OM}}$ improves not only the tunability of the cavity, but also increases the radiation pressure force. Accessing a regime in which radiation pressure becomes significant, either from an external laser source or the internal laser field itself, opens up several new possibilities for all-optical laser wavelength tuning and locking~\cite{ref:Alegre_Perahia,ref:rosenberg_static_2009}. 

% several dominate over thermo-optic effects requires better thermal dissipation.  One possible approach towards this regime involves shrinking the suspended area of the nanobeams at the cost of tunability.  Such a design would also facilitate continuous-wave electrical pumping due to improved heat-sinking and contacting to the substrate.  

This work was supported by the DARPA NACHOS program (Award No. W911NF-07-1-0277).  The authors would like to thank Jianxin Chen for growth of the laser material, and the Kavli Nanoscience Institute at Caltech. 

%\bibliography{./bibliography_RP_6_22_2009}

\begin{thebibliography}{17}
\expandafter\ifx\csname natexlab\endcsname\relax\def\natexlab#1{#1}\fi
\expandafter\ifx\csname bibnamefont\endcsname\relax
  \def\bibnamefont#1{#1}\fi
\expandafter\ifx\csname bibfnamefont\endcsname\relax
  \def\bibfnamefont#1{#1}\fi
\expandafter\ifx\csname citenamefont\endcsname\relax
  \def\citenamefont#1{#1}\fi
\expandafter\ifx\csname url\endcsname\relax
  \def\url#1{\texttt{#1}}\fi
\expandafter\ifx\csname urlprefix\endcsname\relax\def\urlprefix{URL }\fi
\providecommand{\bibinfo}[2]{#2}
\providecommand{\eprint}[2][]{\url{#2}}

\bibitem[{\citenamefont{Lee and Wu}(2005)}]{ref:Lee_MCM}
\bibinfo{author}{\bibfnamefont{M.-C.~M.} \bibnamefont{Lee}} \bibnamefont{and}
  \bibinfo{author}{\bibfnamefont{M.~C.} \bibnamefont{Wu}},
  \bibinfo{journal}{IEEE Photonics Tech. Lett.} \textbf{\bibinfo{volume}{17}},
  \bibinfo{pages}{1034} (\bibinfo{year}{2005}).

\bibitem[{\citenamefont{Huang et~al.}(2008)\citenamefont{Huang, Zhou, and
  {Chang-Hasnain}}}]{ref:Huang_MCY}
\bibinfo{author}{\bibfnamefont{M.~C.~Y.} \bibnamefont{Huang}},
  \bibinfo{author}{\bibfnamefont{Y.}~\bibnamefont{Zhou}}, \bibnamefont{and}
  \bibinfo{author}{\bibfnamefont{C.~J.} \bibnamefont{{Chang-Hasnain}}},
  \bibinfo{journal}{Nature Photonics} \textbf{\bibinfo{volume}{2}},
  \bibinfo{pages}{180} (\bibinfo{year}{2008}).

\bibitem[{\citenamefont{Frank et~al.}(2009)\citenamefont{Frank, Deotare,
  {McCutcheon}, and Lon\v{c}ar}}]{ref:loncar_dynamically_2009}
\bibinfo{author}{\bibfnamefont{I.~W.} \bibnamefont{Frank}},
  \bibinfo{author}{\bibfnamefont{P.~B.} \bibnamefont{Deotare}},
  \bibinfo{author}{\bibfnamefont{M.~W.} \bibnamefont{{McCutcheon}}},
  \bibnamefont{and}
  \bibinfo{author}{\bibfnamefont{M.}~\bibnamefont{Lon\v{c}ar}},
  \bibinfo{journal}{Opt. Express} \textbf{\bibinfo{volume}{18}},
  \bibinfo{pages}{7505} (\bibinfo{year}{2009}).

\bibitem[{\citenamefont{Duarte}(2008)}]{ref:duarte_tunable_2008}
\bibinfo{author}{\bibfnamefont{F.~J.} \bibnamefont{Duarte}},
  \emph{\bibinfo{title}{Tunable Laser Applications, Second Edition}}
  (\bibinfo{publisher}{{CRC}}, \bibinfo{year}{2008}), \bibinfo{edition}{2nd}
  ed., ISBN \bibinfo{isbn}{1420060090}.

\bibitem[{\citenamefont{{Chang-Hasnain}}(2000)}]{ref:Chang_C_J}
\bibinfo{author}{\bibfnamefont{C.~J.} \bibnamefont{{Chang-Hasnain}}},
  \bibinfo{journal}{IEEE J. Sel. Top. Quan. Elec.}
  \textbf{\bibinfo{volume}{6}}, \bibinfo{pages}{978} (\bibinfo{year}{2000}).

\bibitem[{\citenamefont{Kippenberg and
  Vahala}(2008)}]{ref:kippenberg_cavity_2008}
\bibinfo{author}{\bibfnamefont{T.~J.} \bibnamefont{Kippenberg}}
  \bibnamefont{and} \bibinfo{author}{\bibfnamefont{K.~J.}
  \bibnamefont{Vahala}}, \bibinfo{journal}{Science}
  \textbf{\bibinfo{volume}{321}}, \bibinfo{pages}{1172} (\bibinfo{year}{2008}).

\bibitem[{\citenamefont{Povinelli et~al.}(2005)\citenamefont{Povinelli,
  Lon\v{c}ar, Ibanescu, Smythe, Johnson, Capasso, and
  Joannopoulos}}]{ref:Povinelli_evanescent}
\bibinfo{author}{\bibfnamefont{M.~L.} \bibnamefont{Povinelli}},
  \bibinfo{author}{\bibfnamefont{M.}~\bibnamefont{Lon\v{c}ar}},
  \bibinfo{author}{\bibfnamefont{M.}~\bibnamefont{Ibanescu}},
  \bibinfo{author}{\bibfnamefont{E.~J.} \bibnamefont{Smythe}},
  \bibinfo{author}{\bibfnamefont{S.~G.} \bibnamefont{Johnson}},
  \bibinfo{author}{\bibfnamefont{F.}~\bibnamefont{Capasso}}, \bibnamefont{and}
  \bibinfo{author}{\bibfnamefont{J.~D.} \bibnamefont{Joannopoulos}},
  \bibinfo{journal}{Opt. Lett.} \textbf{\bibinfo{volume}{30}},
  \bibinfo{pages}{3042} (\bibinfo{year}{2005}).

\bibitem[{\citenamefont{Li et~al.}(2009)\citenamefont{Li, Pernice, and
  Tang}}]{ref:Tang_cavity}
\bibinfo{author}{\bibfnamefont{M.}~\bibnamefont{Li}},
  \bibinfo{author}{\bibfnamefont{W.~H.~P.} \bibnamefont{Pernice}},
  \bibnamefont{and} \bibinfo{author}{\bibfnamefont{H.~X.} \bibnamefont{Tang}},
  \bibinfo{journal}{Phys. Rev. Lett.} \textbf{\bibinfo{volume}{103}},
  \bibinfo{pages}{3901} (\bibinfo{year}{2009}).

\bibitem[{\citenamefont{Eichenfield et~al.}(2009)\citenamefont{Eichenfield,
  Camacho, Chan, Vahala, and Painter}}]{ref:eichenfield_picogram_2009}
\bibinfo{author}{\bibfnamefont{M.}~\bibnamefont{Eichenfield}},
  \bibinfo{author}{\bibfnamefont{R.}~\bibnamefont{Camacho}},
  \bibinfo{author}{\bibfnamefont{J.}~\bibnamefont{Chan}},
  \bibinfo{author}{\bibfnamefont{K.~J.} \bibnamefont{Vahala}},
  \bibnamefont{and} \bibinfo{author}{\bibfnamefont{O.}~\bibnamefont{Painter}},
  \bibinfo{journal}{Nature} \textbf{\bibinfo{volume}{459}},
  \bibinfo{pages}{550} (\bibinfo{year}{2009}).

\bibitem[{\citenamefont{Deotare et~al.}(2009)\citenamefont{Deotare,
  {McCutcheon}, Frank, Khan, and Lon\v{c}ar}}]{ref:Loncar_zipper}
\bibinfo{author}{\bibfnamefont{P.~B.} \bibnamefont{Deotare}},
  \bibinfo{author}{\bibfnamefont{M.~W.} \bibnamefont{{McCutcheon}}},
  \bibinfo{author}{\bibfnamefont{I.~W.} \bibnamefont{Frank}},
  \bibinfo{author}{\bibfnamefont{M.}~\bibnamefont{Khan}}, \bibnamefont{and}
  \bibinfo{author}{\bibfnamefont{M.}~\bibnamefont{Lon\v{c}ar}},
  \bibinfo{journal}{Appl. Phys. Lett.} \textbf{\bibinfo{volume}{95}},
  \bibinfo{pages}{1102} (\bibinfo{year}{2009}).

\bibitem[{\citenamefont{Alegre et~al.}(2010)\citenamefont{Alegre, Perahia, and
  Painter}}]{ref:Alegre_Perahia}
\bibinfo{author}{\bibfnamefont{T.~P.~M.} \bibnamefont{Alegre}},
  \bibinfo{author}{\bibfnamefont{R.}~\bibnamefont{Perahia}}, \bibnamefont{and}
  \bibinfo{author}{\bibfnamefont{O.}~\bibnamefont{Painter}},
  \bibinfo{journal}{Opt. Express} \textbf{\bibinfo{volume}{18}},
  \bibinfo{pages}{7872} (\bibinfo{year}{2010}).

\bibitem[{\citenamefont{Chan et~al.}(2009)\citenamefont{Chan, Eichenfield,
  Camacho, and Painter}}]{ref:chan_optical_2009}
\bibinfo{author}{\bibfnamefont{J.}~\bibnamefont{Chan}},
  \bibinfo{author}{\bibfnamefont{M.}~\bibnamefont{Eichenfield}},
  \bibinfo{author}{\bibfnamefont{R.}~\bibnamefont{Camacho}}, \bibnamefont{and}
  \bibinfo{author}{\bibfnamefont{O.}~\bibnamefont{Painter}},
  \bibinfo{journal}{Opt. Express} \textbf{\bibinfo{volume}{17}},
  \bibinfo{pages}{3802} (\bibinfo{year}{2009}).

\bibitem[{\citenamefont{Hwang et~al.}(1998)\citenamefont{Hwang, Baillargeon,
  Chu, Sciortino, and Cho}}]{ref:Hwang2}
\bibinfo{author}{\bibfnamefont{W.-Y.} \bibnamefont{Hwang}},
  \bibinfo{author}{\bibfnamefont{J.}~\bibnamefont{Baillargeon}},
  \bibinfo{author}{\bibfnamefont{S.~N.~G.} \bibnamefont{Chu}},
  \bibinfo{author}{\bibfnamefont{P.~F.} \bibnamefont{Sciortino}},
  \bibnamefont{and} \bibinfo{author}{\bibfnamefont{A.~Y.} \bibnamefont{Cho}},
  \bibinfo{journal}{J. Vac. S. Tech. B} \textbf{\bibinfo{volume}{16}},
  \bibinfo{pages}{1422} (\bibinfo{year}{1998}).

\bibitem[{\citenamefont{Perahia et~al.}(2009)\citenamefont{Perahia, Alegre,
  {Safavi-Naeini}, and Painter}}]{ref:Perahia_R}
\bibinfo{author}{\bibfnamefont{R.}~\bibnamefont{Perahia}},
  \bibinfo{author}{\bibfnamefont{T.~P.~M.} \bibnamefont{Alegre}},
  \bibinfo{author}{\bibfnamefont{A.~H.} \bibnamefont{{Safavi-Naeini}}},
  \bibnamefont{and} \bibinfo{author}{\bibfnamefont{O.}~\bibnamefont{Painter}},
  \bibinfo{journal}{Appl. Phys. Lett.} \textbf{\bibinfo{volume}{95}},
  \bibinfo{pages}{201114} (\bibinfo{year}{2009}).

\bibitem[{\citenamefont{Bj$\ddot{\text{o}}$rk
  et~al.}(1992)\citenamefont{Bj$\ddot{\text{o}}$rk, Karlsson, and
  Yamamoto}}]{ref:Bjork4}
\bibinfo{author}{\bibfnamefont{G.}~\bibnamefont{Bj$\ddot{\text{o}}$rk}},
  \bibinfo{author}{\bibfnamefont{A.}~\bibnamefont{Karlsson}}, \bibnamefont{and}
  \bibinfo{author}{\bibfnamefont{Y.}~\bibnamefont{Yamamoto}},
  \bibinfo{journal}{Appl. Phys. Lett.} \textbf{\bibinfo{volume}{60}},
  \bibinfo{pages}{304} (\bibinfo{year}{1992}).

\bibitem[{\citenamefont{Rosenberg et~al.}(2009)\citenamefont{Rosenberg, Lin,
  and Painter}}]{ref:rosenberg_static_2009}
\bibinfo{author}{\bibfnamefont{J.}~\bibnamefont{Rosenberg}},
  \bibinfo{author}{\bibfnamefont{Q.}~\bibnamefont{Lin}}, \bibnamefont{and}
  \bibinfo{author}{\bibfnamefont{O.}~\bibnamefont{Painter}},
  \bibinfo{journal}{Nature Photon.} \textbf{\bibinfo{volume}{3}},
  \bibinfo{pages}{478} (\bibinfo{year}{2009}), ISSN \bibinfo{issn}{1749-4885}.

\bibitem[{\citenamefont{Camacho et~al.}(2009)\citenamefont{Camacho, Chan,
  Eichenfield, and Painter}}]{ref:camacho_characterization_2009}
\bibinfo{author}{\bibfnamefont{R.~M.} \bibnamefont{Camacho}},
  \bibinfo{author}{\bibfnamefont{J.}~\bibnamefont{Chan}},
  \bibinfo{author}{\bibfnamefont{M.}~\bibnamefont{Eichenfield}},
  \bibnamefont{and} \bibinfo{author}{\bibfnamefont{O.}~\bibnamefont{Painter}},
  \bibinfo{journal}{Opt. Express} \textbf{\bibinfo{volume}{17}},
  \bibinfo{pages}{15726} (\bibinfo{year}{2009}).

\end{thebibliography}

\end{document}